\input{aipcheck}
\documentclass[
    ,final            
  ]
  {aipproc}

\layoutstyle{8x11single}
     
\newcommand{\dslash}{{\partial\!\!\!\!\!\!/}}     

\begin{document}

\title{Neutron Stars and the High Density Equation of State}

\classification{12.38.Lg 12.39.Ki 21.65.Qr 26.60.Kp 97.60.Jd}
\keywords      {neutron stars, nuclear matter, quark matter, equation of state, QCD phase transition}

\author{T.~Kl\"ahn}{
  address={Physics Division, Argonne National Laboratory, Argonne IL 60439, USA}
}

\author{C.D.~Roberts}{
  address={Physics Division, Argonne National Laboratory, Argonne IL 60439, USA}
}

\author{D.B.~Blaschke}{
  address={Institute for Theoretical Physics, University
of Wroclaw, 50-204 Wroclaw, Poland}
  ,altaddress={Bogoliubov Laboratory for Theoretical Physics,
JINR Dubna, 141980 Dubna, Russia} 
}

\author{F.~Sandin}{
  address={IFPA, D\'epartement AGO, Universit\'e de Li\`ege, Sart
 Tilman, 4000 Li\`ege, Belgium}
}

\begin{abstract}
One of the key ingredients to understand the properties of
neutrons stars\footnote{We use the generic term 
{\it neutron stars} for simplicity.
In the context of this paper
this does not exclude any exotic form of strongly interacting matter in the NS core such as color
superconducting quark matter~\cite{Blaschke:1999qx}.}
(NS)
 is the equation of state at finite densities 
far beyond nuclear saturation. 
Investigating the phase structure of 
quark matter that might be realized in
the core of NS inspires
theory and observation.
We discuss recent results of our work
to point out our view on challenges and possibilities
in this evolving field by means of a few examples. 
\end{abstract}

\maketitle


\section{Introduction}
Not only due to their compactness with central densities multiples 
beyond nuclear saturation NS are considered to belong
to the most interesting systems nature offers 
to study the properties of ultra-dense matter.
But it is this feature that provides a basis for the
wide range of unique possibilities that NS offer 
to gain insight into the very nature of fundamental interactions.
A fascinating and yet undecided question
is whether the QCD phase transition from nucleonic to the
fundamental quark matter (QM) degrees of freedom is realized
in the core of NS.
Answering this question in a strict sense
would require the theorist to describe nuclear matter
at finite densities in the framework of QCD,
which is challenging already in vacuum.
Solving QCD in medium
at this fundamental level is a yet unsolved task
even though promising work into this
direction has been done.
In a common approach to bypass this problem 
quark and nuclear matter EoS are modeled separately
and are combined afterwards by constructing a
phase transition thermodynamically.
However, already  pure quark or nuclear
matter EoS are barely constrained
due to the few reliable data
at densities beyond saturation.

This paper embraces aspects of our recent
work on ultra-dense matter.
It starts with a discussion
of potential promising constraints on the 
EoS of ultra-dense matter
from NS mass and mass-radius measurements
and the elliptic flow in heavy ion collisions.
These, other constraints and more detailed discussions
can be found in \cite{Klahn:2006ir,Lattimer:2006xb,Blaschke:2008cu}.
The two following sections consider the
EoS in nuclear and quark matter.
Here we focus on the importance of
constraining in particular the nuclear EoS \cite{Klahn:2006ir}
and discuss present approaches on
describing quark matter in medium.
While deconfinement and
dynamical chiral symmetry breaking are
accessible by solving QCD's gap equation
within the framework of the 
Dyson-Schwinger formalism \cite{Chen:2008zr},
effective models of the NJL-type 
are a practical tool to explore the
phase structure of quark matter at
finite densities \cite{Klahn:2006iw,Blaschke:2008br}.
Finally, we discuss a recently introduced approach
to construct the phase transition from nuclear
to quark matter which results in a possibly mixed
phase of one-flavor quark matter on a 
background of nuclear matter \cite{Blaschke:2008br}.

\section{Constraints on the equation of state}

Far from the basic perception of a NS as an object as heavy, and heavier,  
as our sun, bound by gravity into a sphere with a diameter
of about 25~km and spinning around its axis up to 1000 times per second
(which is already fascinating) 
{\it NS are complex and inherently versatile objects}.
This is because
{\it NS have an internal structure and evolve in time.}
These two elements 
- structure and temporal history -
are both a challenge and the key
for a detailed understanding 
of the physics of matter in NS.
With increasing observational efforts
especially over the last decade,
which will increase further over the next,
NS data become more and more valuable
as constraints for the modeling
of ultra-dense matter.
Of particular interest are
NS which significantly deviate 
from the norm since those define the natural limits
one has to account for in any reliable theoretical description.

Facing the present potential  uncertainties in observation and modeling, it makes sense to start from an 'extremal set' of observables 
(largest and smallest known NS mass, radius, moment of inertia,
rotational period, etc.)
and to examine the consistency of this set,
e.g. for similar underlying model assumptions.
The benefit of this approach is apparent in two ways.
First, a reliable EoS should be
in agreement with all trusted NS data available.
Second, taking into account extremal data
is a valuable additional strategy 
to explore the real limits of NS properties and the EoS.
Therefore, at present, the  challenge is to identify 
both reliable ``extreme'' NS data and
possible {\it contradicting or complementing data that might narrow
the limits of NS observables and the EoS}.

\subsection{Neutron Star Masses}

For the purpose of this work we focus on
NS configurations which are extreme in terms of
their mass and/or radius. Due to observational
uncertainties one should be careful not to
mistake these extreme values as the final truth on the subject.
A relativistic analysis of timing observations of the pulsar
PSR J1903+0327 finds its mass to be $1.74\pm04 {\rm M_\odot}$.
This is the largest pulsar mass that has ever been precisely measured \cite{Champion:2008ge}. 
Evidence for more massive NS exists but suffers
from larger uncertainties.
We consider a maximum NS mass of about $2.0 {\rm ~M_\odot}$,
which is in agreement 
with timing measurements for PSR B1516+02B, located in the
globular cluster M5.
The derived pulsar mass is $(2.08\pm 0.19)~{\rm
  M_\odot}$ (at 68\% probability) with a 95\% probability that the mass of
this object is above $1.72~{\rm M_\odot}$ \cite{Freire:2007xg} which deviates
considerably from the average masses of binary radio pulsars, ${\rm M_{BRP}}=1.35\pm
0.04~{\rm M_{\odot}}$ \cite{Thorsett:1998uc}.
However, looking at the recent history and the prominent
example of PSR J0751+1807 we are aware of the
fact that the identification of NS masses is a sensitive task
and might give flawed results.
In the latter case the mass initially determined to be
$2.1\pm0.2 {\rm M_\odot} (1\sigma)$ had to be corrected to
a value of $1.26\pm0.2 {\rm M_\odot} (1\sigma)$ due to
new data analysis.
There are other observations pointing towards high NS masses
($2.10\pm0.28~{\rm M_\odot}$ for EXO 0748-676,$2.0\pm0.1~{\rm M_\odot}$ for  4U 1636-536) but for the given reason 
they should be treated with caution.
As we will point out, the analysis of the elliptic flow 
in heavy ion collisions suggests
that $2 {\rm ~M_\odot}$ is an approximate
upper limit on NS masses, but presently no other criterion
is known, that would decisively rule out the existence of high massive NS. 
It will depend on future observations 
to affirmatively determine the actual upper 
limit on NS masses.

\subsection{Radii and Mass-Radius Relations}

\begin{figure}[htb] 
\includegraphics[keepaspectratio,height=0.5\textwidth,
angle=-90]{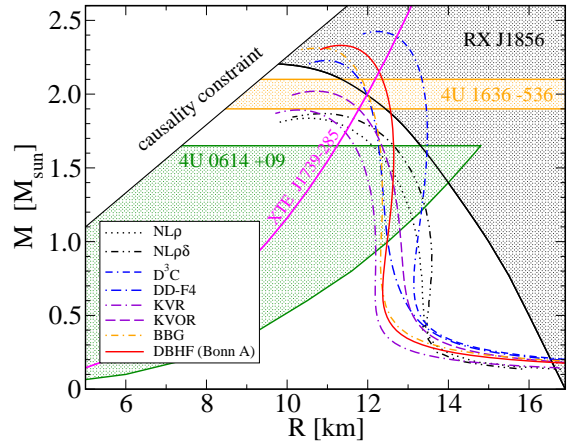} 
\caption{Mass-radius constraints from thermal radiation of the
  isolated NS RX J1856 (grey hatched region) and from QPOs in the
  LMXBs 4U 0614+09 (wedge-like green-hatched area) and 4U 1636-536 (orange
  hatched region).  For 4U 1636-536 a mass of $2.0\pm 0.1~{\rm M_\odot}$ is obtained
  so that the weak QPO constraint would exclude the NL$\rho$ and
  NL$\rho\delta$ EoS, whereas the strong one renders only the stiffest EoS
  D$^3$C, DD-F4, BBG and DBHF viable. 
  XTE J1739-285 \cite{Kaaret:2006gr,Lattimer:2006xb}
  would favor a soft EoS,
  which is controversial, however (see text). 
}
\label{fig:M-R-nuclear} 
\end{figure}

Aside from NS masses, kilohertz quasi-periodic brightness
oscillations (QPOs) seen from more than 25 low-mass NS X-ray binaries (LMXBs)
can be used to put additional constraints on the high-density EoS. A pair of
such QPOs is often seen from these systems \cite{vanderKlis:2000ca}. In all currently
viable models for these QPOs, the higher QPO frequency is close to the orbital
frequency at some special radius. For such a QPO to last the required many
cycles (up to $\sim 1000$ in some sources), the orbit must be outside the
star. According to general relativity theory the orbit must also be outside
the innermost stable circular orbit (ISCO).  Gas or particles inside the ISCO
would spiral rapidly into the star, preventing the production of sharp QPOs.
This implies \cite{Miller:2003wa,Miller:1996vj} that the observation of a
source whose maximum QPO frequency is $\nu_{\rm max}\approx\nu_{\rm ISCO}$ limits the stellar mass
and radius to
\begin{equation} 
\begin{array}{rl} 
   M < 2.2~{\rm M_\odot} {{1000~{\rm Hz}} \over {\nu_{\rm max}}} 
(1+0.75j) ~ , \quad
  R < 19.5~{\rm km}
{{1000~{\rm Hz}} \over {\nu_{\rm max}}}
(1+0.2j) ~ .
\label{eq:qpo}
\end{array} 
\end{equation} 
The quantity $j\equiv cJ/GM^2$ (with $J$ the stellar angular momentum) is the
dimensionless spin parameter, which is typically in the range between 0.10 and
0.2 for these systems. Equation (\ref{eq:qpo}) implies that for given observed
value $\nu_{\rm max}$ the mass and radius of that source must be inside a
wedge-shaped area, as shown in Fig.\ \ref{fig:M-R-nuclear}. Since the wedge becomes smaller for higher $\nu_{\rm max}$, the highest frequency ever
observed, 1330~Hz for 4U~0614+091 \cite{vanStraaten:2000gd}, places the strongest
constraint on the EoS.  As can be seen from Fig.\ \ref{fig:M-R-nuclear},
the current QPO constraints do not rule out any of the EoS considered
here. However, because higher frequencies imply smaller wedges, the future
observation of a QPO with a frequency in the range of $\sim 1500-1600$~Hz
would rule out the stiffest of our EoS.
 
If there is evidence for a particular source that a given frequency is
close to the orbital frequency at the ISCO, then the mass is known to
a good accuracy, with uncertainties arising from the spin parameter.
This was first claimed for 4U~1820--30 \cite{Zhang:1998di}, but
complexities in the source phenomenology have made this controversial.
More recently, careful analysis of Rossi X-ray Timing Explorer data
for 4U~1636--536 and other sources \cite{Barret:2005wd} has suggested
that sharp and reproducible changes in QPO properties are related to
the ISCO.  If so, this implies that several NS in low-mass
X-ray binaries should have gravitational masses between $1.9\,{\rm M_\odot}$
and possibly $2.1\,{\rm M_\odot}$ \cite{Barret:2005wd}. In Fig.\
\ref{fig:M-R-nuclear} we show the estimated mass of $2.0\pm
0.1\,{\rm M_\odot}$ for 4U~1636--536.

Recently, mass-radius constraints have been reported for the accreting
compact stars XTE J1739-285 \cite{Lavagetto:2006ew} and SAX
J1808.4-3658 \cite{Leahy:2007fb} (SAX J1808 for short) which are based
on the identification of the burst oscillation frequency with the spin frequency of the compact star.  
It is almost impossible to fulfill the
constraints from RX J1856 (and other high-mass candidates) and SAX
J1808 (which favors a soft EoS) simultaneously so that the status of SAX J1808 is currently controversial. It is likely that the small radius estimate of Leahy et al. \cite{Leahy:2007fb} is a consequence of the underestimation of higher harmonics when only timing data are analyzed and not also the energy spectra. 

The nearby isolated NS RX J1856.5-3754 (RX J1856 for short) belongs to a group of seven objects which show a purely thermal spectrum in X-rays and in optical-UV.  
This allows the determination of $R_\infty/d$, the ratio of
the photospheric radius $R_\infty$ to the distance $d$ of the object, if the
radiative properties of its photosphere are known.  RX J1856 is the only
object of this group which has a measured distance obtained by Hubble Space
Telescope (HST) astrometry.  After the distance of 117~pc \cite{Walter:2002uq}
became known several groups pointed out that the blackbody radius of this star
is as large as 15 to 17~km.  Although both the X-ray and the optical-UV
spectra are extremely well represented by blackbody functions they require
different emission areas, a smaller hot spot and a larger cooler region.  The
overall spectrum could also be fitted by blackbody emission from a surface
showing a continuous temperature distribution between a hot pole and a cool
equator, as expected for a magnetized NS. The resulting blackbody
radii are 17~km (two blackbodies) and 16.8~km (continuous temperature
distribution) \cite{Trumper:2003we}. In this paper we adopt the result of the
continuous temperature fit, $R_\infty=16.8$ km.  More recent HST observations
of RX J1856 indicated larger distances of up to 178~pc. A distance of around
140~pc for RX J1856 is considered a conservative lower
limit.
For a distance of 140~pc the corresponding radius is
17~km \cite{Ho:2006}. Although some questions--in particular that of the
distance--are not yet finally settled, the recent data support an unusually
large radius for RX J1856.5-3754.

\subsection{Elliptic Flow in Heavy Ion Collisions}

\begin{figure}[t]
\label{fig:flow_extrapol_mass}
\includegraphics[width=0.42\textwidth, angle=-90]{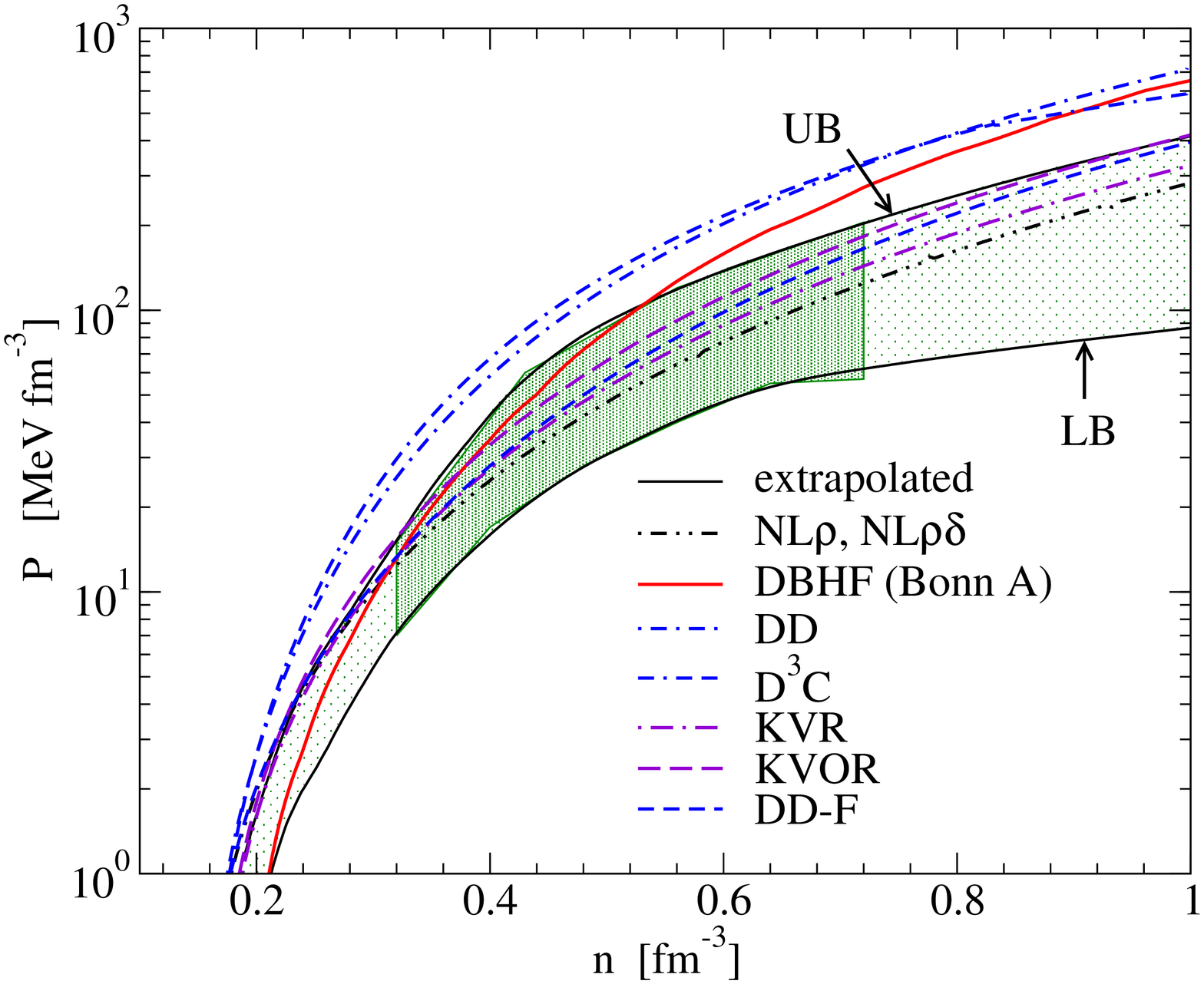}
\hfill
\includegraphics[width=0.42\textwidth, angle=-90]{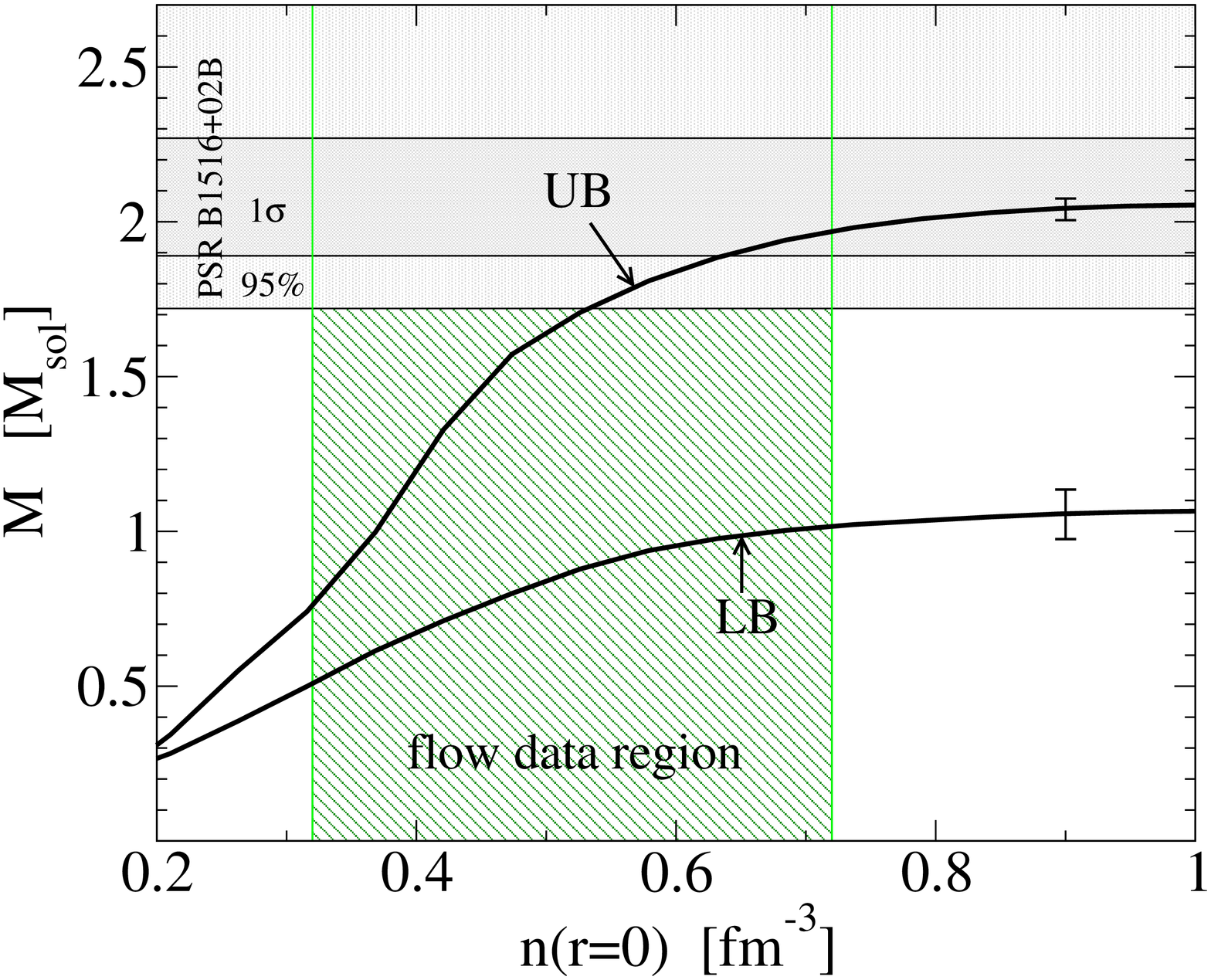}
\caption{
Left panel: Pressure region consistent with experimental flow data
in symmetric nuclear matter (dark shaded region).
The light shaded region extrapolates this region to higher densities
within an upper (UB) and lower border (LB).
Right panel: Mass versus central density for compact star configurations,
calculated using the UB and LB extrapolations of the flow constraint
boundaries.
The error bars take into account
different models for the symmetry energies.
The gray {horizontal} bars denote the expected mass of PSR B1516+02B
within the $1\sigma$ interval and  a likelihood of a  mass below $1.72 {\rm M_\odot}$
of $5\%$, resp.
The vertical bars limit the density region
covered by the flow constraint.}
\end{figure}

The flow of matter in heavy ion collisions
is directed both forward and perpendicular (transverse)
to the beam axis.
At high densities spectator nucleons may shield
the transversal flow into their direction and generate
an inhomogeneous density and thus a pressure profile
in the transversal plane.
This effect is commonly referred to as
elliptic flow and depends directly on the given EoS.
An analysis of these nucleon flow data, 
which depends essentially only on the isospin independent part of the EoS, 
was carried out in a particular model in Ref.~\cite{Danielewicz:2002pu}. 
In particular it was determined for which range of parameters 
of the EoS the model is still compatible with the flow data. 
The region thus determined is shown in  Fig.~{\ref{fig:flow_extrapol_mass}} 
as the dark shaded region.
Ref.~\cite{Danielewicz:2002pu} then asserts that this region
limits the range of accessible pressure values at a given density.
For our purposes we extrapolated this region
by an upper (UB) and lower (LB) boundary,
enclosing the light shade region in Fig.~{\ref{fig:flow_extrapol_mass}}.

Thus the area of allowed values does not represent experimental values itself,
but results from transport calculations
for the motion of nucleons in a collision \cite{Danielewicz:2002pu}.
Of course, it would be preferable to 
use the actual flow data as a
constraint for each specific EoS,
but a corresponding testing tool 
based on a transport code is not 
yet available.
Therefore we adopt the results of ref.~\cite{Danielewicz:2002pu} as a
reasonable estimate of the preferable pressure-density domain in symmetric nuclear matter.
Its upper boundary is expected to be stable against temperature
variations \cite{PrivCom:Danielewicz2006}.
The important fact is that the flow constraint
probes essentially only the symmetric part of 
the energy per nucleon $E_0(n)$.
Following Ref. \cite{Danielewicz:2002pu} the constraint arises for a density window 
between 2 and 4.5  times saturation density $n_s$. 
One has, however, to keep in mind that at high
densities this constraint is based on flow data from the AGS
energy regime ($E_{\rm lab} \sim 0.4 - 11$ AGeV). At these energies
a large amount of the initial bombarding energy is converted into
new degrees of freedom, i.e. excitations of higher lying
baryon resonances and mesons, which makes conclusions on the
nuclear EoS more ambiguous than at low energies. 
Nevertheless, the analysis of \cite{Danielewicz:2002pu} provides 
a guideline for the high density
regime which we believe to be reasonable.
The uncertainties 
are represented by a region
of possible pressures in Fig.~\ref{fig:flow_extrapol_mass}.

An EoS which is in accordance with
the flow constraint might still violate the maximum NS mass constraint.
This is demonstrated in Fig.~\ref{fig:flow_extrapol_mass},
where in particular the lower  boundary of
the limiting region in Fig.~\ref{fig:flow_extrapol_mass}
was extrapolated to construct an artificial EoS (LB)
in order to obtain the mass-density relation for 
corresponding compact star configurations .
The resulting mass curve in Fig.~\ref{fig:flow_extrapol_mass}
is far from reaching even typical NS mass values as the well-measured mass
$M_{B1913+16}=1.4408 \pm 0.0003~{\rm M_\odot}$ \cite{St04}.
This demonstrates that the lower bound (LB) in
Fig.~\ref{fig:flow_extrapol_mass}
does not agree with astrophysical observations,
which therefore potentially narrow the band of the
flow constraint.

The maximum mass constraint demands
a certain stiffness of  $E_0(n)$  in order to obtain sufficiently large maximum NS masses.
The small influence of $E_S(n)$ on the NS mass
can also be well recognized on Fig.~\ref{fig:flow_extrapol_mass}.
Two different symmetry energies,
necessary to describe NS matter,
were taken from the investigated EoS.
They were chosen in accordance with
the DU-constraint and gave the largest (DD-F)
and smallest (D$^3$C) contribution to the binding energy at $n=1$ fm$^{-3}$.
The resulting deviations of the NS masses are shown 
as error bars on the curves in  Fig.~\ref{fig:flow_extrapol_mass}.
It results in a maximum difference of less than $0.2~{\rm M_\odot}$ for
the mass curves corresponding to LB.
The same was done for an artificial EoS extrapolating the upper boundary (UB).
Here, the largest error estimate of approximately $0.1~{\rm  M_\odot}$ is even smaller.
The maximum mass for UB of about $2.0~{\rm M_\odot}$ again fulfills
the corresponding constraint.

\section{Nuclear Matter}
While modern nuclear matter models give a rather
similar description of the saturation and sub-saturation
behavior, they differ considerably in their extrapolations
to densities above $\approx 2n_S$, the regime which is relevant
for NS physics and heavy ion collisions.
This is illustrated in Table \ref{table:neos}, where
several nuclear EoS are characterized
by comparing the parameters of an expansion 
around saturation as a function of the density deviation $\epsilon = (n-n_s)/
n_s$ and the asymmetry $\beta$ 
according to
\begin{equation}
 E = a_{V} + \frac{K}{18} \epsilon^{2} - \frac{K^{\prime}}{162} \epsilon^{3} +
\dots + \beta^{2} \left( J + \frac{L}{3} \epsilon + \dots \right).
\end{equation}
\begin{table}[t]
\label{table:neos}
\caption{A set of nuclear equations of state. The entries are: saturation density, $n_s$; binding energy, $a_{V}$;
    incompressibility, $K$; skewness parameter, $K^{\prime}$; symmetry energy,
    $J$; symmetry energy derivative, $L$; Dirac effective mass, $m_{D}$.}
\begin{tabular}{@{}l|llrrrrr@{}}
  EoS           & $n_s$         & $a_{V}$   & $K$     & $K^{\prime}$ & $J$   & $L$   & $m_{D}$\\
                & [fm$^{-3}$]   & [MeV]     & [MeV]   & [MeV]        &[MeV]  & [MeV]  & [$m$]\\
 \hline
 NL$\rho$       & $0.1459$      & $-16.062$ & $203.3$ & $ 576.5$     &$30.8$ & $83.1$ & $0.603$ \\
 NL$\rho\delta$ & $0.1459$      & $-16.062$ & $203.3$ & $ 576.5$     &$31.0$ & $92.3$ & $0.603$ \\
 D${}^{3}$C     & $0.1510$      & $-15.981$ & $232.5$ & $-716.8$     &$31.9$ & $59.3$ & $0.541$ \\
 DD-F4           & $0.1469$     & $-16.028$ & $220.4$ & $ 1229.2$     &$32.7$ & $58.7$ & $0.556$ \\
 KVR            & $0.1600$      & $-15.800$ & $250.0$ & $ 528.8$     &$28.8$ & $55.8$ & $0.805$ \\
 KVOR           & $0.1600$      & $-16.000$ & $275.0$ & $ 422.8$     &$32.9$ & $73.6$ & $0.800$ \\
 DBHF           & $0.1810$      & $-16.150$ & $230.0$ & $ 507.9$     &$34.4$ & $69.4$ & $0.678$ \\
 BBG            & $0.1901$      & $-14.692$ & $221.6$ & $ -132.4$    &$36.3$ & $79.4$ & $-$        \\
 DD-RH          & $0.172$       & $-15.73$  & $249.0$ & $-$          &$34.4$ & $90.2$ & $0.686$ \\
\end{tabular}
\end{table}
Most of the shown models are derived in the framework of the
relativistic mean-field approach,\cite{Wal74,Ser86,Rei89,Rin96} allowing for
non-linear (NL) self-interactions of the $\sigma$ meson
\cite{Gaitanos:2003zg}.  For model NL$\rho$, the isovector part of the
interaction is described entirely in terms of $\rho$ meson exchange. This is
different for NL$\rho\delta$ where the isovector part of the interaction is
described in terms of both $\rho$ and $\delta$-meson exchange. The latter is
generally neglected in RMF models \cite{Liu:2001iz}. RMF models with density
dependent input parameters (coupling constants and masses) are represented in
Table \ref{table:neos} by four different models from two classes, where in the first one
density dependent meson couplings are modeled such that several properties of
finite nuclei (binding energies, charge and diffraction radii, surface
thicknesses, neutron skin in ${}^{208}$Pb, spin-orbit splittings) can be
fitted \cite{Typel:2005ba}. D${}^{3}$C additionally contains a derivative
coupling which leads to momentum-dependent nucleon self-energies, and DD-F4 is
modeled such that the flow constraint from heavy ion collisions is
fulfilled \cite{Danielewicz:2002pu}. The second class of these models is
motivated by the Brown-Rho scaling assumption \cite{Brown:1991kk} that not only
the nucleon mass but also the meson masses should decrease with increasing
density.  In the KVR and KVOR models \cite{Kolomeitsev:2004ff} these
dependences are related to a nonlinear scaling function of the $\sigma$-meson
field such that the EoS of symmetric nuclear matter and pure neutron matter
below four times the saturation density coincide with those of the
Urbana-Argonne group \cite{Akmal:1998cf}. In this way the latter approach builds a
bridge between the phenomenological RMF models and a microscopic EoS built on
realistic nucleon-nucleon forces. The RMF models are contrasted with several
variational models for the EoS such as APR\cite{Akmal:1998cf},
WFF\cite{Wiringa:1988tp}, FPS\cite{Friedman:1981qw}, a relativistic
Dirac-\-Brueckner-\-Hartree-\-Fock (DBHF) model\cite{vanDalen:2004pn,vanDalen:2005ns}, and a
nonrelativistic Brueckner-\-Bethe-\-Goldstone (BBG) model \cite{Baldo:1999rq}.

\begin{figure}[htb]
  \includegraphics[keepaspectratio,height=0.60\textwidth,
  angle=-90]{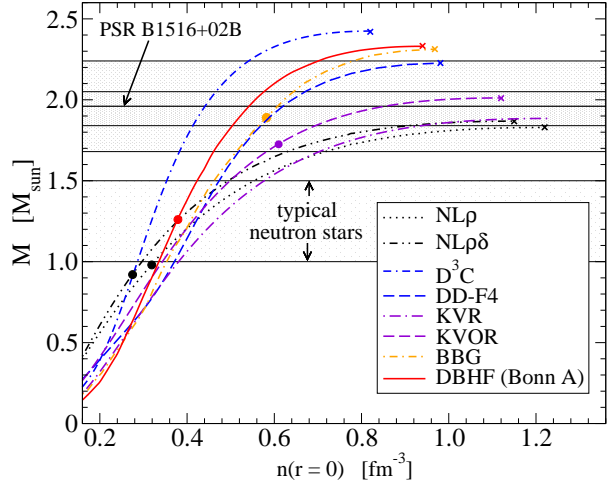}
  \caption{Mass versus central density of NS, 
    computed for the EoS shown in Table~\ref{table:neos}.  Crosses refer to
    the maximum-mass star of each sequence, filled dots mark the critical
    masses and central densities beyond which the direct Urca (DU) cooling
    process becomes possible.}
  \label{fig:M-n} 
\end{figure} 

As shown in Fig.\ \ref{fig:M-n} none of these EoS predicts a
maximum NS mass below
the $2\sigma$ mass limit of $1.68 ~{\rm M_\odot}$ for PSR B1516+02B, and even at the
$1\sigma$ mass limit of $1.84 ~{\rm M_\odot}$ the softest EoS NL$\rho$ and
NL$\rho\delta$ cannot be excluded.  
This underlines the important rule of the maximum NS mass as a constraint on the high density EoS.
If a pulsar with a mass
exceeding $1.8-1.9~{\rm M_\odot}$ at the 2$\sigma$ or even 3$\sigma$ level would be
observed in the future, this will impose a severe constraint on the stiffness
of the nuclear EoS. For the set of EoS tested here, only the stiffest 
models, i.e. D$^3$C, DD-F4, BBG, and DBHF would remain viable candidates.
It is noteworthy, that the flow constraint limits the
stiffness of the EoS to be 'rather' soft, 
whereas the higher the constraining maximum NS mass is
the 'stiffer' the EoS needs to be. Both together provide
complementing data.

\section{Quark Matter}
\subsection{Nonperturbative QCD}
Confinement and dynamical chiral symmetry breaking (DCSB) are key emergent phenomena in QCD.  
Neither of these is apparent in QCD's Lagrangian and 
yet they play a dominant role in determining the 
observable characteristics of real-world QCD.
A natural starting point for an exploration of DCSB is QCD's gap equation. 
\begin{equation}
 S(p;\mu)^{-1}= Z_2 (i\vec{\gamma}\cdot \vec{p} + i \gamma_4 (p_4+i\mu) + m^{\rm bm}) + \Sigma(p;\mu)\,, \label{gendse}
\end{equation}
with the renormalized self energy expressed as
\begin{eqnarray}
\Sigma(p;\mu ) &=& Z_1 \int^\Lambda_q\! g^2(\mu) D_{\rho\sigma}(p-q;\mu) 
\frac{\lambda^a}{2}\gamma_\rho S(q;\mu) \Gamma^a_\sigma(q,p;\mu) , \label{gensigma}
\end{eqnarray}
where $\int^\Lambda_q$ represents a translationally invariant regularization of the integral, with $\Lambda$ the regularization mass-scale, $D_{\rho\sigma}(k;\mu)$ is the dressed-gluon propagator, $\Gamma^a_\sigma(q,p;\mu)$ is the dressed-quark-gluon vertex, and $m^{\rm bm}$ is the $\Lambda$-dependent current-quark bare mass.  The quark-gluon-vertex and quark wave function renormalization constants, $Z_{1,2}(\zeta^2,\Lambda^2)$, depend on the renormalization point, $\zeta$, the regularization mass-scale and the gauge parameter.  A nonzero chemical potential introduces no divergences in addition to those present in the $\mu=0$ theory.  Hence, the renormalization constants determined at $\mu=0$ are completely sufficient in-medium.
The gap equation's solution can assume the general form \cite{Rusnak:1995ex}
\begin{eqnarray} 
S(p;\mu)^{-1} = i \vec{\gamma}\cdot \vec{p} \, A(p^2,p\cdot u,\zeta^2)
+  i \gamma_4(p_4+i\mu) \, C(p^2,p\cdot u,\zeta^2) + B(p^2,p\cdot u,\zeta^2)  \,,
%
\label{sinvp} 
\end{eqnarray}
where  $u=(\vec{0},i\mu)$. The mass function $M(p^2,p\cdot u)= B(p^2,p\cdot u,\zeta^2)/ A(p^2,p\cdot u,\zeta^2)$ is renormalization point independent.
There are numerous studies of this basic Dyson-Schwinger equation 
in-vacuum \cite{Maris:2003vk,Roberts:2007jh} and in-medium \cite{Roberts:2000aa}. 
It is noteworthy that extant self-consistent studies of concrete models of QCD,
which exhibit both confinement and DCSB in vacuum, 
possess coincident deconfinement and chiral symmetry restoration transitions; 
e.g., Refs.\,\cite{Bender:1996bm,Blaschke:1997bj,Bender:1997jf,Maris:2001rq,Bashir:2008fk}. 
This result appears to follow from the crucial role played by the 
in-medium evolution of the dressed-quark self-energy in both transitions.

Significant effort continues to be expended on determining the precise nature of 
the kernel of QCD's gap equation.  
A dialogue between DSE studies and results 
from numerical simulations of lattice-regularized QCD is providing important information; 
e.g., Refs.\,\cite{Bhagwat:2003vw,Alkofer:2003jj,Bhagwat:2004hn,Bhagwat:2004kj,Bhagwat:2006tu,%
Kizilersu:2006et,Kamleh:2007ud,Boucaud:2008ky,Cucchieri:2008fc}.  
This body of work can be used to formulate reasonable \textit{Ans\"atze}
for the dressed-gluon propagator and dressed-quark-gluon vertex in Eq.\,(\ref{gensigma}).

DCSB is signaled by the appearance of a gap equation solution 
in which the Dirac-scalar piece is nonzero in the chiral limit.  
This effect owes primarily to a dense cloud of gluons that clothes 
a low-momentum quark \cite{Bhagwat:2007vx,Roberts:2007ji}.  
DCSB is the single most important mass generating mechanism 
for light-quark hadrons; e.g., one can identify it as being 
responsible for roughly 98\% of a proton's mass.  
A system in which the gap equation's DCSB solution is 
the favored ground state is said to realize chiral symmetry 
in the Nambu-Goldstone mode.  
The antithesis is termed a Wigner-Weyl realization of the symmetry.  
Since a system's ground state is that configuration for which 
the pressure is a global maximum 
the pressure difference between the Nambu-Goldstone and Wigner
phase reveals which of both is realized.
\begin{figure}[t]
\includegraphics[scale=0.810] {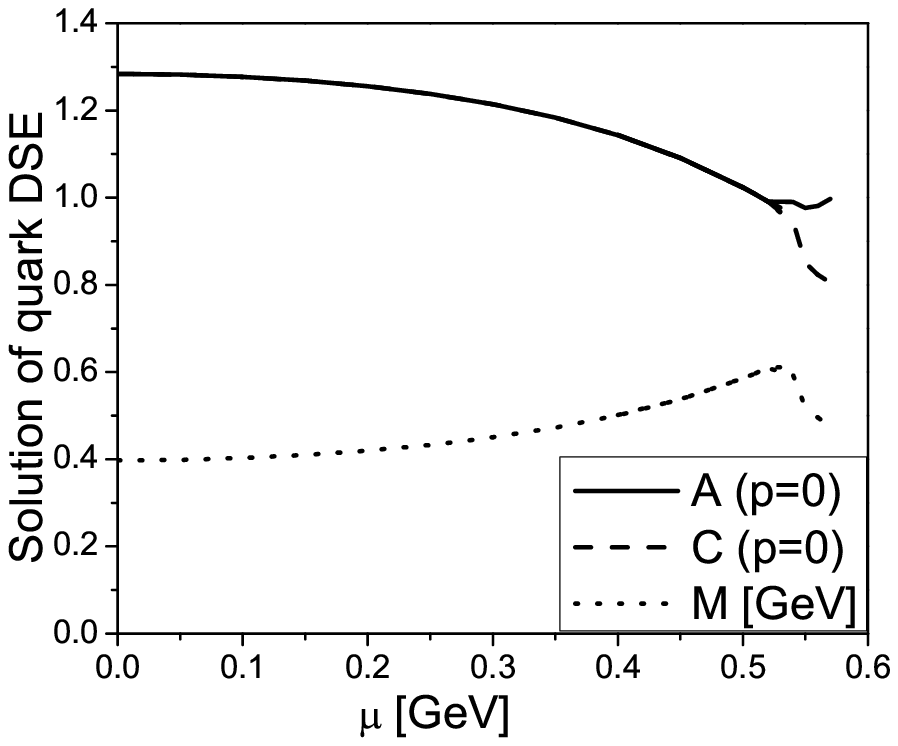} 
\includegraphics[scale=0.810] {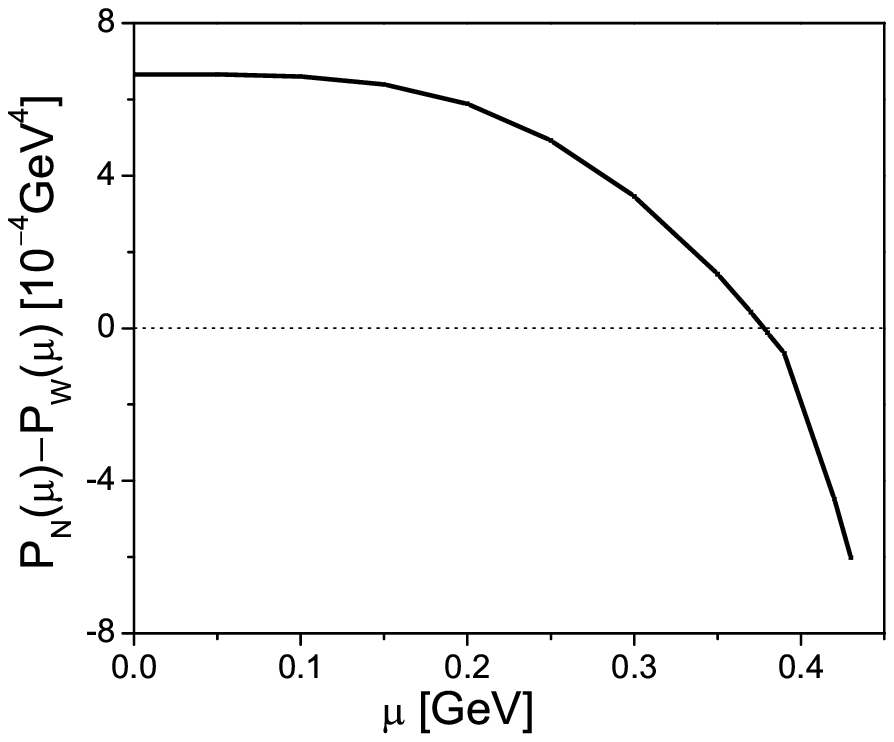} 
\caption{Left panel: Evolution with $\mu$ of the dimensionless 
quantities $A(p=0;\mu)$, $C(p=0;\mu)$, and $M(p=0;\mu)$ measured in GeV, 
computed from the gap equation (\ref{gensigma}).
Right Panel: Difference in pressure 
between the Nambu-Goldstone and Wigner solutions of the gap equation.
Observe, that the Wigner phase is realized for $\mu>0.38$ GeV,
which results in a first order phase DCSB transition at this value.
\label{fig:DCSB}
}
\end{figure}
A striking signature of dynamical chiral symmetry breaking 
is a nonzero value of the dressed-quark mass function $M(p,\mu)$, 
which entails a nonzero chiral-limit quark condensate.  
Naturally, a nonzero chemical potential destabilizes
the quark condensate and, with increasing chemical potential, 
results in a first-order chiral symmetry restoring transition 
at $\mu \approx M(0)$. This is illustrated in Fig. \ref{fig:DCSB}.
Evidence  suggests a coincident first-order 
deconfinement transition \cite{Chen:2008zr}.

\subsection{Effective Quark Matter Models}
There is no inherent obstacle that prohibits an extension of
the powerful Dyson-Schwinger approach
to the finite density domain 
and in fact it appears as the most promising candidate 
for a future systematic modeling of dense QCD matter in compact stars \cite{Roberts:2000aa,Blaschke:1997bj,Nickel:2006kc}.
However, for explorative studies on the possibly rich phase
structure of dense matter in QCD it is for practical reasons
useful to employ effective models which preserve certain but
not necessarily all features of QCD.
Increasingly applied are models of the  Nambu--Jona-Lasinio (NJL)
type where chiral symmetry is dynamically broken in the 
nonperturbative vacuum and restored at finite densities and temperatures.
Although the standard NJL model is quite simple,  
based on a few parameters, such as momentum cutoff and 
coupling constants for a set of interaction channels, 
it offers possibilities for generalizations,  
like a density-dependence of the cutoff parameter \cite{Baldo:2006bt},
momentum-dependent formfactors \cite{Schmidt:1994di} 
and an infrared cutoff  to mimic 
confinement  \cite{Blaschke:1998ws,Lawley:2006ps}.  

Essentially one starts from a {\it QCD-motivated} partition function
\begin{eqnarray}   
\label{Z}  
Z(T,\hat{\mu})&=&\int {\mathcal D}\bar{q}{\mathcal D}q   
\exp \left\{\int_0^\beta d\tau\int d^3x\,\left[   
        \bar{q}\left(i\dslash-\hat{m}_0+\hat{\mu}\gamma^0\right)q+  
{\mathcal L}_{\rm int}   
\right]\right\},  
\end{eqnarray} 
which explicitly accounts for phases of possible
interest by considering specific channels in 
the interaction part ${\mathcal L}_{\rm int}$
of the Lagrangian.
Mass gaps, pairing gaps and the EoS are then obtained from the 
mean-field thermodynamic potential, $\Omega_{MF}=-T\ln Z_{MF}$.
As example serves the following model \cite{Blaschke:2008vh} 
with diquark pairing 
in the single flavor color-spin-locking (CSL), 
two-flavor (2SC) and three-flavor 
color-flavor locking (CFL) channels,
\begin{eqnarray}   
\label{Lint}  
{\mathcal L}_{\rm int} &=& G_S\left\{ \sum_{a=0}^{8}
        \left[(\bar{q}\tau_aq)^2 + (\bar{q}i\gamma_5\tau_a q)^2 \right]
        +\eta_{D0}\hspace{-3mm}\sum_{A=2,5,7} j_{D0,A}^\dagger  j_{D0,A} 
        +\eta_{D1}~j_{D1}^\dagger j_{D1} \right\},  \nonumber 
\end{eqnarray}
where 
$\hat{\mu}=\frac{1}{3}\mu_B
+{\rm diag}_f(\frac{2}{3},-\frac{1}{3},-\frac{1}{3})\mu_Q
+\lambda_3\mu_3+\lambda_8\mu_8$ 
is the diagonal quark chemical  potential matrix  
and $\hat{m}_0={\rm diag}_f(m_u^0,m_d^0,m_s^0)$  
the current-quark mass matrix.  
The relative coupling strengths of the 
spin-0 and spin-1 diquark currents, 
$j_{D0,A}=q^TiC\gamma_5\tau_A\lambda_Aq$ and  
$j_{D1}=q^TiC(\gamma_1\lambda_7+\gamma_2\lambda_5+\gamma_3\lambda_2)q$, 
are essentially free parameters,  but can be chosen according to 
a Fierz transformation of the one-gluon exchange 
interaction, $\eta_{D0}=3/4$ and $\eta_{D1}=3/8$, see  \cite{Aguilera:2005tg}. 
We point out, that bosonised diquark fields
in QCD play an essential role in the
color singlet baryon states \cite{Cahill:1988bh}.

The next section shows
how observational constraints help to fix a model EoS
obtained in this  framework.

\section{the quark hadron phase transition}

\begin{figure} [t]
\includegraphics[angle=270,width=0.5\textwidth]{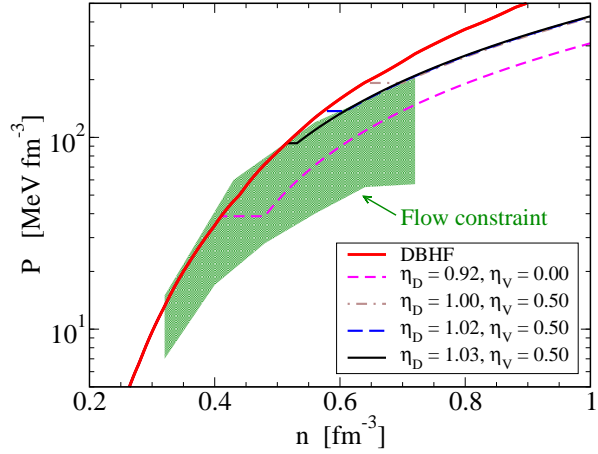}
\caption{Hybrid EoS  with a low density hadronic 
branch described by the DBHF approach and a high density quark matter branch 
obtained from a three-flavor NJL model with color superconductivity (diquark 
coupling $\eta_D$) and isoscalar vector meson coupling ($\eta_V$).}
    \label{fig:P-n}
\end{figure}
A strategy to construct hybrid EoS in
agreement with observational data is to
adjust available coupling strengths
within certain limits to 
a subset of available data.
This is demonstrated in Fig.\ref{fig:P-n} for a model with an interaction 
different from that in Eq.(\ref{Lint}).
Here the coupling strength of the scalar diquark ($\eta_D$) and 
isoscalar vector meson ($\eta_V$)  channel
are considered as basically free parameters
and have been adjusted so, that after constructing
the (Maxwell-)phase transition from the nuclear DBHF EoS
to this QM EoS the hybrid EoS fits the upper limit of
the flow constraint \cite{Blaschke:2007fr}.
Just by adjusting to this requirement
the hybrid EoS fulfills all 
recently developed constraints from modern compact star   
observations  \cite{Klahn:2006ir}.
It is an interesting and encouraging result,
that none of the present constraints   
on the mass-radius relation of NS rules out hybrid stars \cite{Alford:2006vz,Klahn:2006iw}.

\begin{figure}[t]   
\includegraphics[angle=0,height=0.46\textwidth]{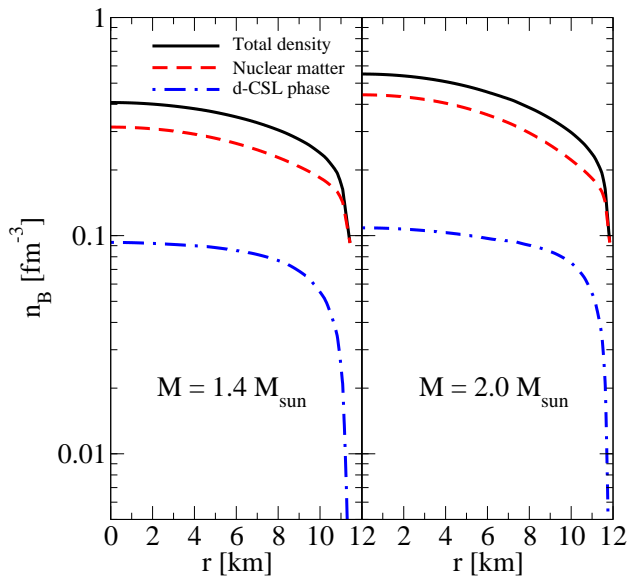}  
\caption{   
Density profiles of two stars with masses $1.4$~M$_\odot$ and    
$2.0$~M$_\odot$. 
{In this model a mixed phase of d-CSL quark 
matter with nuclear matter extends up to the crust-core boundary.} 
\label{fig:profile}}  
\end{figure}

As a last point we discuss ambiguities
in the way  the phase transition is modeled.
The $\beta$-equilibrium condition in compact stars 
relates the chemical potentials of quarks and electrons by 
$\mu_d=\mu_s$ and $\mu_d=\mu_u+\mu_e$. 
The mass difference between the strange and the light quark flavors  
$m_s > m_u, m_d$ induces different densities of the down and strange 
quark. Charge neutrality requires a 
finite electron density and consequently  $\mu_d>\mu_u$. 
When increasing the baryochemical potential, 
the d-quark chemical potential 
is the first to reach a critical value where the  
the partial density of free d-quarks becomes finite 
in a first order phase transition. 
Due to the finite value of $\mu_e$ the u-quark chemical potential  
is below the critical value and the s-quark density is zero due to 
the high s-quark mass. 
Therefore the negative electric charge of the d-quark
cannot be compensated in a pure quark phase
and {\it single-flavor} d-quark matter under 
NS conditions does not exit.
The situation changes considering a 
chemical equilibrium of the type 
$n+n \leftrightarrow p + 3 d$,
with positive charged protons
which compensate the charge of 
a {\it single-flavor}  d-quark phase
and results in a mixed phase of nucleons 
and down quarks once the d-quark chemical 
potential exceeds the critical value.
This can be thought analogous to 
the dissociation of nuclear clusters in the 
crust of NS (neutron dripline).
One interesting consequence of the effect
is illustrated in Fig.\ref{fig:profile}.
Ignoring in a first approach important effects of the 
inhomogeneous phase mixture, 
e.g. Coulomb interactions and surface tension,
the existence of such a phase suggests 
small fractions of quark matter even near the 
the crust-core boundary.
We point out, that neither
way to model the phase transition
can be considered to be ``correct''.
Nucleons actually consist of quarks
and as expressed before it
will take serious efforts to describe
the QCD phase transition within an
appropriate and fundamental approach.




\begin{theacknowledgments}
We acknowledge useful communications
with and the support of
E.~F.~Brown, H.~Chen, I.C.~Clo\"et, P.~Danielewicz, 
Y.-X.~Liu, M.C.~Miller, and R.~Rutledge.
We are grateful to the organizers of the 
{\it Fifth ANL/INT/MSU/JINA FRIB Theory Workshop}.
This work was supported by: the Department of Energy, Office of Nuclear Physics, contract no.\ DE-AC02-06CH11357 (TK,CDR);
the Belgian fund for scientific research FNRS (FS);
the Polish Ministry for Research and Higher Education, grant no.\ N N 202 0953 33 (DBB);  
CompStar, an ESF Research Networking Programme (DBB,FS). 
\end{theacknowledgments}



\bibliographystyle{aipproc}   

\bibliography{myrefs}

\IfFileExists{\jobname.bbl}{}
 {\typeout{}
  \typeout{******************************************}
  \typeout{** Please run "bibtex \jobname" to obtain}
  \typeout{** the bibliography and then re-run LaTeX}
  \typeout{** twice to fix the references!}
  \typeout{******************************************}
  \typeout{}
 }


\end{document}